\documentclass[namedreferences]{kluwer}

\newcommand{\cago}{$^{12}{\rm C}(\alpha,\gamma)^{16}{\rm O}$\ }

\usepackage{graphicx}

\begin{document}
\begin{article}

\begin{opening}         

\title{Seismic Inference using Genetic Algorithms}

\author{Travis S. \surname{Metcalfe}}
\institute{Theoretical Astrophysics Center, Aarhus University, Denmark}

\runningauthor{T.S. Metcalfe}
\runningtitle{Seismic Inference using Genetic Algorithms}

\begin{abstract}
A flood of reliable seismic data will soon arrive. The migration to larger
telescopes on the ground may free up 4-m class instruments for multi-site
campaigns, and several forthcoming satellite missions promise to yield
nearly uninterrupted long-term coverage of many pulsating stars. We will
then face the challenge of determining the fundamental properties of these
stars from the data, by trying to match them with the output of our
computer models. The traditional approach to this task is to make informed
guesses for each of the model parameters, and then adjust them iteratively
until an adequate match is found. The trouble is: how do we know that our
solution is unique, or that some other combination of parameters will not
do even better? Computers are now sufficiently powerful and inexpensive
that we can produce large grids of models and simply compare {\it all} of
them to the observations. The question then becomes: what range of
parameters do we want to consider, and how many models do we want to
calculate? This can minimize the subjective nature of the process, but it
may not be the most efficient approach and it may give us a false sense of
security that the final result is {\it correct}, when it is really just
{\it optimal}. I discuss these issues in the context of recent advances in
the asteroseismological analysis of white dwarf stars.
\end{abstract}


\end{opening}

\section{Wampler's Screwdriver}

Most scientists are familiar with the concept of Occam's razor---the idea
that if you have to choose between competing explanations for some
physical phenomenon, the simplest explanation is most likely to be
correct. My thesis supervisor, Ed Nather, told me a story about another
less widely known scientific tool that may be just as important as Occam's
razor. He calls it ``Wampler's screwdriver'' \cite{nat95}.

In the early 1970's, Ed was attending a conference of the Astronomical
Society of the Pacific in California, and Joe Wampler was giving a
presentation about the first discovery of a double quasar \cite{wam73}.
The standard procedure at the time was to identify blue objects inside the
relatively large positional error box of a newly discovered radio point
source, and then take spectra of them, one by one, until you found one
with a big redshift. What Joe decided to do was go back to the fields
where quasars had been discovered in this way, and take spectra of {\it
all} of the blue objects, even after he found one of them to be a quasar.
Joe's double quasar turned out to be an accidental alignment of two
quasars at different distances, but later on others repeated what he had
done and found a double quasar that was the result of gravitational
lensing---so his method was an important contribution to the field. At the
end of Joe's presentation, he posed a simple question to the audience:
``Why do you always find a lost screwdriver in the last place you look?''
The answer, of course, is because you stop looking.

In this paper, I will review a method of fitting models to seismological
data that {\it keeps looking}, even after it has found a pretty good fit
to the observations. This is drawn from work I have been doing over the
past few years to develop a model-fitting method based on a genetic
algorithm, helping us to learn more about pulsating white dwarf stars
(\opencite{mnw00}; \citeyear{mwc01}). In section~\ref{wdsec}, I will
discuss what this method has allowed us to learn about white dwarfs, but
for the bulk of this review I will focus on the method and related issues.

\begin{figure}
\centerline{\includegraphics[width=12truecm]{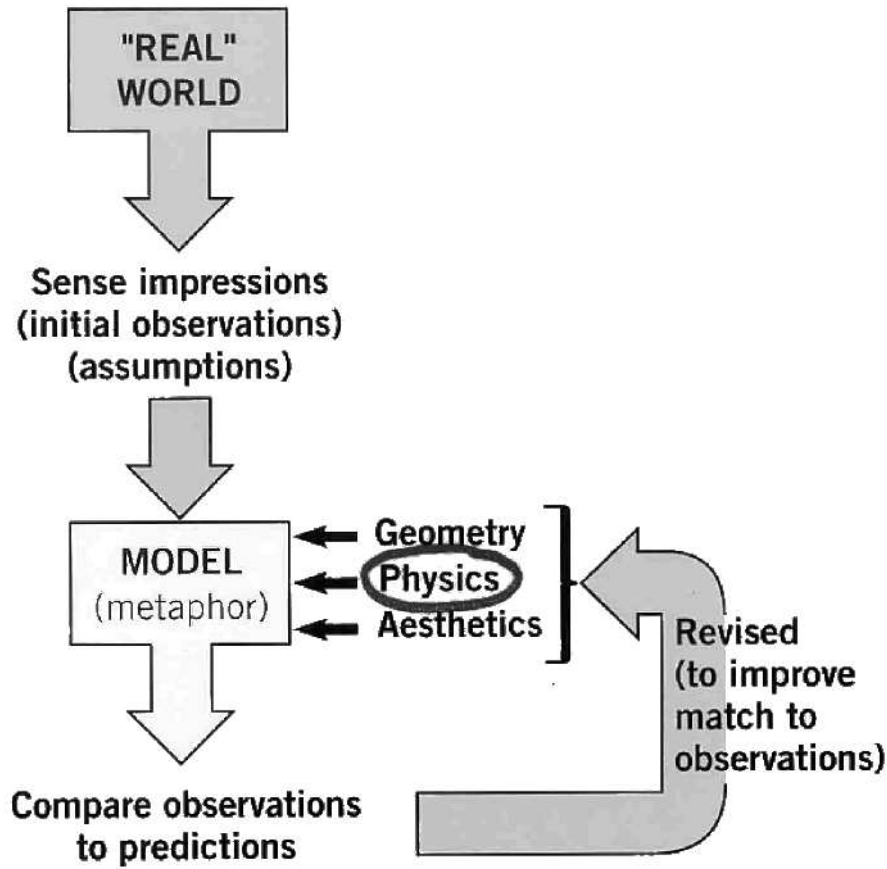}}
\caption[]{A representation of the logical distinction between reality, 
observations, and models of reality. Note that the ultimate goal of this
process is an improved understanding of the constitutive physics.}
\label{fig1}
\end{figure}

\subsection{Reality, Models, \& Physics}

Before I get to that, I would like to step back and take a philosophical
look at the general process that we use to learn anything about the
objects we study (see Figure~\ref{fig1}). Out there somewhere there is a
``real world'' that we pass through various filters and selection effects
to get ``observations''. As observers, we try our best to compensate for
every effect between the real world and the data point, but it's important
to realize that we use models in this process too. 

With the observations in hand, we devise computer models to try to explain
them, doing our best to include all of the relevant physical processes
that are in principle detectable. Generally these models have a number of
tunable parameters, and we do our best to adjust them until the
predictions of the model agree as closely as possible with the
observations---in our case, generally the pulsation periods of a star.
When we have found a model that adequately reproduces the observations, we
assume that the values of the parameters tell us something about the
properties of the actual star. But we should never forget that what we are
actually dealing with are {\it models} of reality, and not reality itself.
When we derive values of $\log g$ and $T_{\rm eff}$ from spectral lines,
for example, we are not measuring the mass and temperature of the
star---we are (at best) deriving the optimal match between our models of
stellar atmospheres and the extracted spectrum over a finite wavelength
interval. We call this approach the ``forward method'', and it can only
tell us what is best {\it within the context of the models we use}. It
cannot tell us that we are using the wrong models, unless or until we
actually try different models.

\begin{figure}
\centerline{\includegraphics[width=12truecm]{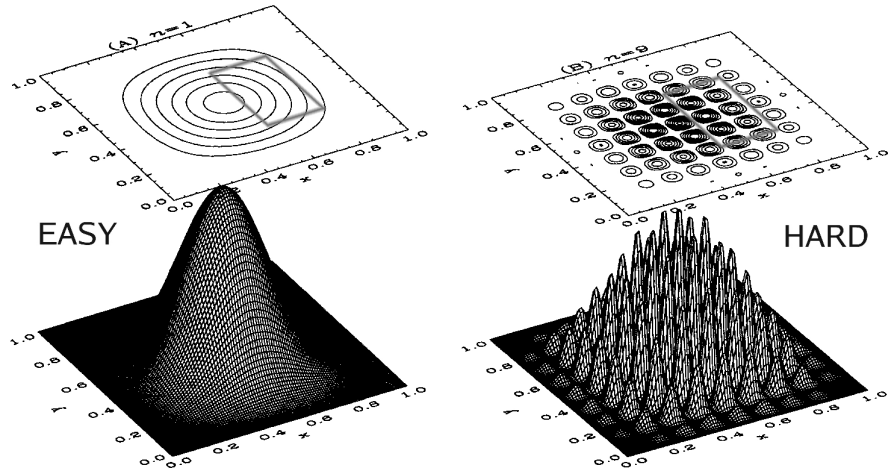}}
\caption[]{Example model-spaces that would be easy (left) and hard (right) 
for traditional optimization methods. Note how the likely outcome in 
the two cases would differ if the search were confined to the region 
specified by the grey rectangle in the contour plots (Adapted from 
\opencite{cha95}).} 
\label{fig2}
\end{figure}

\section{Optimization \& Objectivity \label{optsec}}

With this caveat in mind, the process of trying to adjust our model
parameters to fit the observations is really just an optimization problem.  
Mathematicians have devised various methods, each with their strengths and
weaknesses, to approach such problems. Imagine that the two axes of the
plots shown in Figure~\ref{fig2} represent the two parameters of a model,
and that the height of the surface for each combination of parameters is
some measure of how well the model matches the observations. If the
surface looks like the plot on the left, then just about any optimization
method will work. But if the surface looks more like the plot on the
right, then most traditional optimization methods will fail miserably,
most likely ending up on top of one of the smaller peaks---effectively
finding a locally optimal fit, rather than the globally optimal solution.
Of course, the trouble is that in general we do not {\it know} what the
shape of this surface will be for a given model unless we actually
evaluate it at each of these points. And that's exactly what we want to
avoid by using an optimization method in the first place: we want the
solution as quickly as possible, but we also want it to be the global
solution. In section~\ref{gasec}, I will demonstrate how a genetic
algorithm offers a nice tradeoff between these two competing demands.

Before even choosing an optimization method, there is a more fundamental
way that we can bias our final result: by defining the range of our search
too narrowly. Once again, if the model space is simple, as in the left
side of Figure~\ref{fig2}, we will probably get a final result at the edge
of our search range if we have defined it too narrowly. But if the model
space is more complicated, we can find a ``global'' solution well inside
our search range that is not really globally optimal. For this reason, it
is important that we define the limits of our search as broadly as
possible---constrained only by the physics of the model, and by
observations.

For the models of pulsating white dwarfs that I have been using, for
example, we adjust five different parameters: the stellar mass, the
effective temperature, the mass of the surface helium layer, and two
parameters to describe the internal carbon/oxygen profile. Our search
range includes masses between 0.45 and 0.95 $M_{\odot}$: white dwarfs with
lower masses are expected to have a helium core, and almost all white
dwarfs with known masses are below our upper limit \cite{nap99}. For now
we have been concentrating on the simplest, helium-atmosphere (DB) white
dwarfs, so we allow temperatures between 20,000 and 30,000 K, which easily
encompasses the spectroscopic temperatures of all known pulsators whether
or not trace amounts of hydrogen are included in the atmospheres
\cite{bea99}. We allow the surface helium layer mass to be anywhere from
10$^{-2}~M_*$ (a larger mass would theoretically lead to nuclear burning
at the base of the layer) down to a few times 10$^{-8}$, close to the
limit where our models no longer pulsate \cite{bw94a}. We allow the
internal composition to range from pure carbon to pure oxygen, and we use
the fifth parameter to specify the fractional mass location where the
composition begins to change from a uniform C/O mixture in the center,
which is what we expect from evolutionary models \cite{sal97}. The
location of this composition transition is expected to be reasonably close
to the half-mass point of the model, but we allow it to be anywhere from
0.1 to 0.85 in fractional mass.

\section{Model Fitting}

Once we have defined the search range, the only 100\% reliable method of
finding the globally optimal set of model parameters for a given set of
observations is to calculate an entire grid. The main thing to decide at
this point is the appropriate resolution of the grid in each parameter.  
It is common to do this by deciding how long we want to wait for the
computer to finish, and which parameters are the most interesting. But
even if we had infinite computing power, it makes no sense to calculate a
grid so fine that adjacent points give fits that differ by less than the
observational noise. A good way to quantify the uncertainties on observed
pulsation periods is to identify all of the combination frequencies in the
power spectrum, and see how much they differ from what is expected based
on the measured parent frequencies.

When we did this for a pulsating DB white dwarf, we found deviations of a
few hundredths of a second on pulsation periods that were between 400-800
seconds \cite{mnw00}. This implied that it made sense to calculate about
100 points along each of the five parameters within the ranges specified
above. The full grid at this resolution would require 10$^{10}$ model
evaluations, which would take more than a year to finish even if we had
1000 of today's fastest processors. If we wanted this kind of resolution
for just 1, 2, or possibly 3 of the five model parameters, then maybe a
full grid would make sense, especially if we could reuse it for
observations of many objects. But certainly for problems with a higher
number of free parameters, and to avoid recalculating the grid every few
years when the physics gets updated significantly, genetic algorithms can
provide almost everything that a grid search can, at a small fraction of
the computational cost.

\subsection{Genetic Algorithms \label{gasec}}

The basic idea behind a genetic algorithm is fairly simple: it is just an
iterative Monte Carlo method that samples the model space randomly, but
keeps a sort of memory of what worked well in the past. It accomplishes
this through a computational analogy with the idea of biological evolution
through natural selection. The model parameters serve as the genetic
building blocks, and the observations provide the selection pressure. It
starts just like a simple Monte Carlo, where we generate $N$ random sets
of parameters, evaluate the model for each set, and then compare them to
the observations. The genetic algorithm treats each set of parameters as
an individual in a population, and assigns each a ``fitness'' based on how
well it matches the observations. Next, it selects from this population at
random, with the fittest individuals more likely to survive. Here comes
the weird part: it encodes the parameters into simple strings of numbers,
sort of like chromosomes; it pairs them up and performs operations that
are analogous to breeding and mutation, and then decodes the strings back
into numerical values for the parameters. Now we have a new population, so
we evaluate the model for each case again, and continue the whole process
until the population converges to one region of the model space.

\begin{figure}
\centerline{\includegraphics[width=12truecm]{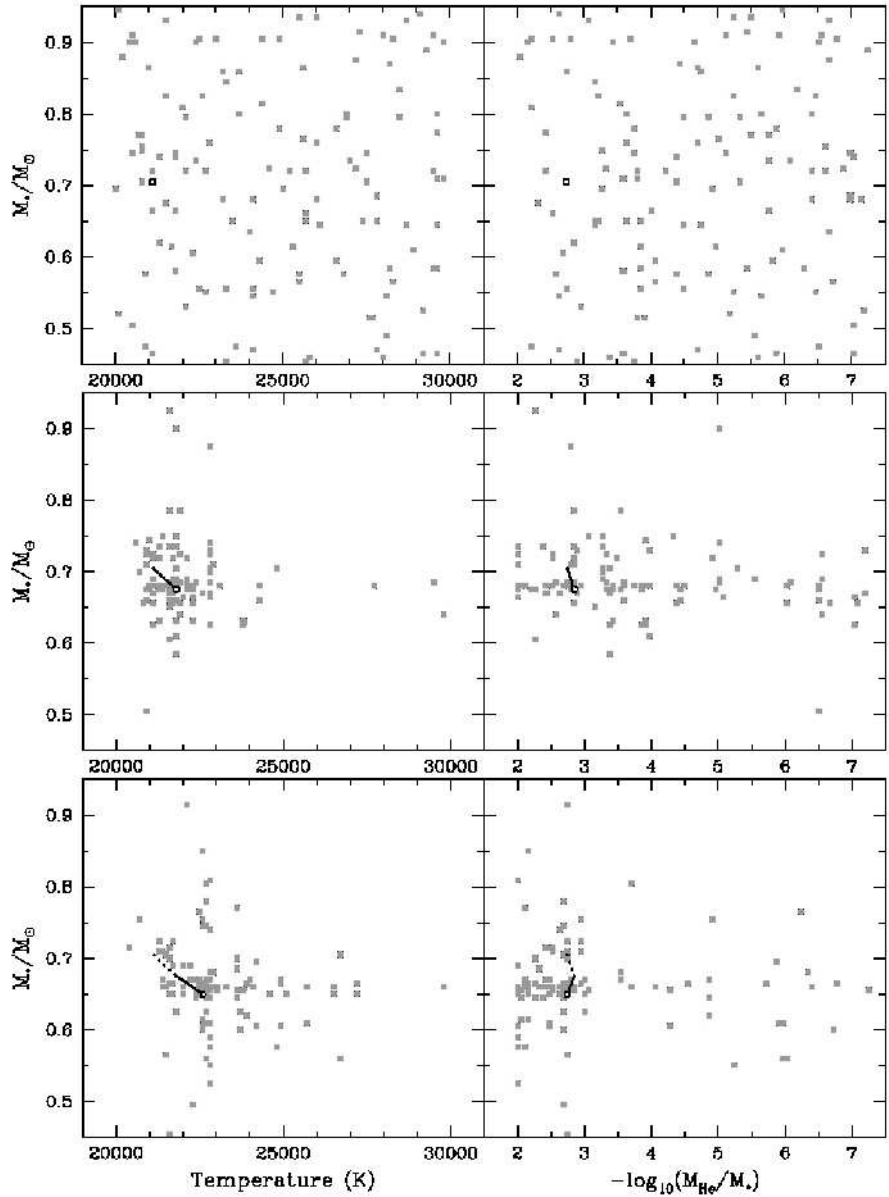}}
\caption[]{Front and side views of the white dwarf model-space, with 
snapshots for generation 0 (top), 80 (middle), and 200 (bottom). The
best parameter-set in each case is indicated by a black open square, 
and yields root-mean-square residuals of 4.0, 1.8, and 1.5 seconds in 
the three panels respectively. The initial convergence is reasonably fast,
but the genetic algorithm {\it keeps looking} for a better solution, 
and eventually finds the globally optimal set of model parameters.}
\label{fig3}
\end{figure}

This is a bit abstract, so let me illustrate how it works in practice,
using a 3-parameter example from the white dwarf problem. The top panel of
Figure~\ref{fig3} shows the initial random sample of the model space:  
each point in the left side of the plot corresponds one-to-one with a
point in the right side---this is a front and side view of the
model-space. There are 128 points in the sample, and the best one is shown
as a black open square. Initially the best root-mean-square (rms)
difference between the observed and calculated pulsation periods is 4
seconds, but after 80 iterations (or ``generations'' of the genetic
algorithm) the sample is starting to narrow in on one region of the
model-space, and the best solution is substantially better than anything
in the original sample. At this point, if we were to compare the observed
and calculated periods, we would judge the fit to be pretty good. But
as I mentioned at the beginning, in the context of Wampler's Screwdriver:
the genetic algorithm {\it keeps looking}. After 200 generations, it has
found the globally optimal solution within this modeling framework.

\begin{figure}
\centerline{\includegraphics[width=12truecm]{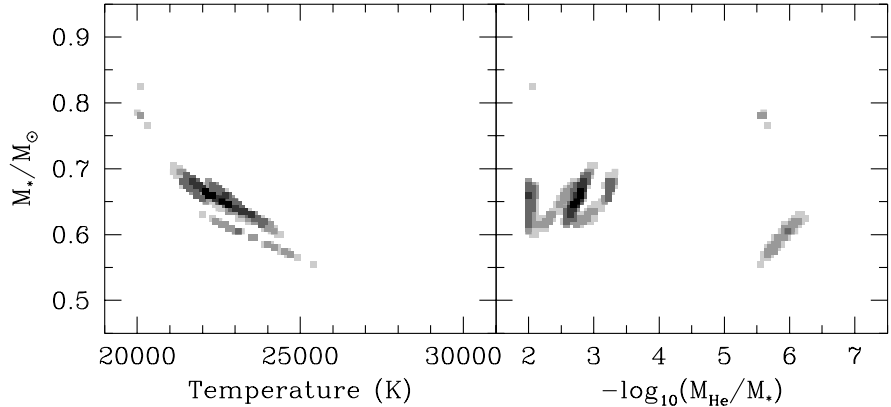}}
\caption[]{Front and side views of the 3-parameter white dwarf
model-space, showing every model evaluated by the genetic algorithm with
rms residuals smaller than 3 seconds. The darkness of each point indicates
the relative quality of the fit. Note the existence of a second family of 
solutions with thin helium layer masses.}
\label{fig4}
\end{figure}

It is worth noting that along the way to the final solution, the genetic
algorithm has evaluated quite a few models, and most of them fall around
better-than-average areas of model-space. So, in the end we also get a
fairly detailed map giving some sense of the uniqueness of the final
solution (see Figure~\ref{fig4}). In this case we see that our optimal
solution has a relatively thick surface helium layer, but there is also a
family of models with thin helium layers that do not match the
observations as well, but still match them better than average. This may
be telling us something about unmodeled structure in the real star
\cite{dk95,bf02}, or possibly about the limitations of the models we are
using \cite{mon02}.

\subsection{Hare \& Hound}

One question that I have not answered yet is: how do we decide when to
stop the genetic algorithm? The answer is what is usually called a ``Hare
\& Hound'' exercise. If you search the Internet for ``Hare \& Hound'', you
will probably find some rather gruesome photographs of a pack of hunting
dogs chasing after a little rabbit and collectively ripping it apart when
they catch it. I do not like the implication of such an image, but I will
use the term anyway. What is usually meant by ``Hare \& Hound'' is a blind
test of the analysis method, where you pass simulated data through a
process to see how well (and how often) you can recover the input data.

Continuing with our example from the 3-parameter application to white
dwarf models, we used the model to calculate the pulsation periods of a
theoretical white dwarf with reasonable values for the mass, temperature,
and surface helium layer mass. Then we picked out the pulsation periods
corresponding roughly to those we had observed in a real white dwarf and
used the genetic algorithm to try to find the model parameters that most
closely reproduced those periods. Since the genetic algorithm relies on
many random processes, it is inevitable that in a finite number of
generations it will sometimes fail to find the input model parameters from
this test. To quantify the success rate, we repeated this experiment 20
times with different random number sequences \cite{mnw00}. In each case,
we saw a rapid improvement after the initial sample, followed by a series
of incremental improvements brought about by random processes. Out of the
20 tests, 9 found the exact input parameters within 200 generations, and
an additional 4 finished with parameters that were close enough to the
input model that a small grid, 11 points on a side, would reveal the exact
values. The other 7 runs got stuck in local minima. So, for any given run
we have a success probability of the method---genetic algorithm plus small
grid---of about 65\%. If we only run it once, there is a 35\% chance
that we will not find the globally optimal solution. But by running it
several times, we gradually increase the probability that we will find the
globally optimal solution, though we pay for this with more model
evaluations. If we do five independent runs, there is a 99.5\% chance that
we will succeed at least once.

\subsection{Problem Scale}

The five independent runs, each with 200 generations of 128 trial
parameter sets, required that we compute the same number of models as a
grid with half the resolution in each dimension. The lesson here is that
for small enough problems, the genetic algorithm is not much more
efficient than a grid. However, when we scale up to 4 or more parameters,
it becomes {\it much} more efficient \cite{mwc01}.  So our hare and hound
test established two things: (1) it told us how many generations we should
let the genetic algorithm run, and (2) it implied that there is some
minimum problem size that should be attempted with the genetic
algorithm---for smaller problems, a model grid might be more efficient.

As an interesting aside, there is also a maximum problem size that should
ever be attempted, and this is set by the rate at which computing power
increases. There is a well known empirical relation called Moore's Law,
which notes that computing power roughly doubles every 18 months---this
has been true since the 1960's. Because of this trend, if a problem is big
enough you can actually get it done faster by waiting for the computing
power to increase before you start it. There was a very humorous paper
published on astro-ph a couple of years ago on this topic \cite{got99},
but the basic conclusions were sound. The authors found that any problem
requiring more than 26 months on the fastest computer presently available
should not be attempted until the future. As a rule of thumb for the
genetic algorithm, a model requiring about 5 minutes to run on today's
fastest available processor will not finish within this 26 month limit on 
a single machine.

\subsection{Parallel Computing}

Even if the problem were able to finish in say 24 months, few of us would
be willing to wait for two years to get the final answer anyway.  
Fortunately, we do not have to confine ourselves to one of today's fastest
processors: we can use many of them. This is one of the great things
about genetic algorithms---or even model grids for that matter: they are
inherently parallelizable. We need to calculate many models, and each of
them is independent of the others. So the number of available processors
sets the number of models that we can evaluate in parallel. Also, there is
very little communication overhead in this process: we send parameter
values out to each processor, and we get back either a list of pulsation
periods, or just a goodness of fit measure if the periods have already
been compared to the observations.

As part of my PhD thesis, I built a 64-processor Linux cluster at the
University of Texas \cite{mn00}. At the time, dual processor machines were
much more expensive than single-processor systems, so it was actually less
expensive to have a separate box for each processor. Now, because
dual-processor systems are much cheaper, and because computing power has
more than quadrupled, we have built a new system that exceeds the speed of
the old system in a much smaller space. We can scale this new system up to
many more units if we need the computing power---but our problems just
don't demand it yet.

In the future, I would guess that stability and space considerations will
be the most important factors for machines like this, and the so-called
``bladed Beowulf'' concept has already reached the point where it is now
possible to put 24 1-GHz processors---none of which require a fan---into a
small cabinet the size of a single desktop computer case \cite{fww02}. The
speed of these processors is upgradeable with improved {\it software},
written as part of his day job by Linus Torvalds, the creator of the Linux
operating system. The big market for these machines, of course, is for
massive web servers---but as a nice side-effect they will also provide
cheap parallel processing for scientists.

\section{Application to White Dwarfs \label{wdsec}}

To give you a sense of the potential of this method, let me summarize what
this has done for white dwarf model-fitting. Allowing five adjustable
parameters (as outlined in section~\ref{optsec}), the genetic algorithm
can find the optimal core composition with a precision of a few percent.
This is a very significant result in itself, because the core composition
can affect the derived ages of white dwarfs by up to 3 Gyr \cite{fbb01},
which has implications for using white dwarfs as chronometers to date
stellar populations. But the central C/O ratio in a white dwarf formed
through single-star evolution can also lead to a measurement of the
important \cago nuclear reaction rate.

When a white dwarf is being formed in the core of a red giant star during
helium burning, the \cago and 3$\alpha$ reactions compete for the
available helium nuclei. As a result, the final central C/O ratio is
primarily determined by the relative rates of these two reactions. The
3$\alpha$ rate is relatively well determined, but the \cago reaction is
much harder to measure, and so its rate is quite uncertain---leading to a
broad range of expected central C/O ratios. By applying the genetic
algorithm method, we can now {\it measure} the central C/O ratio. By
combining this result with evolutionary models that produce internal
chemical profiles, we can tune the \cago rate for a model with a given
mass until we match the derived central C/O ratio \cite{msw02}.

We have now applied this method to observations of two pulsating white
dwarfs, and in the latest results they both yield a reaction rate
consistent with the value derived from laboratory measurements, though
they are at best still marginally consistent with each other \cite{met02}.
We still have more work to do, and crucial tests will come as we apply the
method to additional white dwarfs as observations become available. But we
have come a long way from the subjective, local model-fitting methods of
the past.

\section{Conclusions \& Discussion}

I hope that I have convinced you that genetic algorithms are potentially a
very powerful tool for asteroseismology. They can provide objective global
optimization for problems with more than a few parameters, and in the
process they yield fairly detailed maps of the model-space, which we can
use to judge the uniqueness of the final result. To convince yourself that
a genetic algorithm is more efficient than a large grid, and to convince
others that the final result can be trusted, you should definitely perform
a ``Hare \& Hound'' exercise---trying to match your models to the
observables from another model, and demonstrating that it can be done.  
Remember that the speed of your computer effectively determines the size
of the problems you can attack. If you want to solve a specific problem,
you can and should determine how much computing power you will need to
solve it in a reasonable time. Finally, the payoff can be quite high, as 
it was when we applied this method to our white dwarf models.

I have seen this method begin to transform white dwarf asteroseismology
from a field where the theory was being driven by the observations, to one
where new observations are being driven by the theory. I hope that it can
help launch revolutions in the seismological analysis of other types of
stars too.

\end{article}
\end{document}